\documentclass[]{aa}  
\usepackage{graphicx}
\usepackage{xcolor}
\usepackage{amsmath}
\usepackage{nccmath}
\usepackage{epstopdf}
\usepackage{natbib}
\usepackage{subfig}
\usepackage{txfonts}
\usepackage{hyperref}
\begin{document}

   \title{Evolution of Stokes $V$ area asymmetry related to a quiet Sun cancelation observed with GRIS/IFU}


   \author{Anjali. J. Kaithakkal\inst{1},        
                J. M. Borrero\inst{1}, C. E. Fischer\inst{1},
                C. Dominguez-Tagle\inst{2,3}, and M. Collados \inst{2,3}
                   	      }

   \institute{Leibniz Institute for Solar Physics, Sch\"oneckstrasse 6, 79104 Freiburg, Germany\\
                                {kaithakkal@leibniz-kis.de}         \and
                 Instituto de Astrof\'{i}sica de Canarias, 38205 La Laguna, Tenerife, Spain
				\and
                Departamento de Astrof\'{i}sica, Universidad de La Laguna, 38205 La Laguna, Tenerife, Spain\\                    
             }
\titlerunning{Evolution of Stokes $V$ area asymmetry related to a quiet Sun cancelation }
\authorrunning{Kaithakkal et al.}

   \date{Received ; accepted}

 
  \abstract
{A quiet Sun magnetic flux cancelation event at the disk centre was recorded using the Integral Field Unit (IFU) mounted on the GREGOR Infrared Spectrograph (GRIS). GRIS sampled the event in the photospheric \ion{Si}{I} 10827~{\AA} spectral line. The cancelation is preceded by a significant rise in line core intensity and excitation temperature, which is inferred from Stokes inversions under local thermodynamic equilibrium (LTE). The opposite polarity features seem to undergo reconnection above the photosphere. We also found that the border pixels neighboring the polarity inversion line of one of the polarities exhibit a systematic variation of area asymmetry. Area asymmetry peaks right after the line core intensity enhancement and gradually declines thereafter. Analyzing Stokes profiles recorded from either side of the polarity inversion line could therefore potentially provide additional information on the reconnection process related to magnetic flux cancelation. Further analysis without assuming LTE will be required to fully characterize this event.
 }

  \keywords{Sun: magnetic fields -- Sun: photosphere}

  \maketitle


\section{Introduction}
Magnetic flux cancelation -- the collision of opposite polarity magnetic elements resulting in flux removal from the solar surface -- is a common event \citep[]{1985AuJPh..38..929M, 1985AuJPh..38..855L}. If the opposite polarity features were pre-connected, cancelation involves the retraction of a pre-existing $\Omega$ loop. But if the opposite polarity features were previously unconnected, their cancelation is considered as an observational signature of magnetic reconnection \citep{1987ARA&A..25...83Z}. A number of previous studies have looked at line core intensity enhancement, Doppler velocity values, linear polarization signals, etc., as indicators of reconnection and the associated cancelation \citep[e.g., see][]{2004ApJ...602L..65C, 2005ApJ...626L.125B, 2019A&A...622A.200K}. Another interesting aspect is the investigation of the response of Stokes profiles to the aforementioned dynamic processes. 

Area asymmetry in Stokes $V$ profiles arises from gradients along the line-of-sight (LOS) of the magnetic field and plasma velocity \citep[e.g., ][]{1992ApJ...398..359S}. Using $HINODE$ \citep[]{2007SoPh..243....3K} spectropolarimetric (SP) \citep[]{2001ASPC..236...33L, 2008SoPh..249..167T} observations, \cite{2011A&A...526A..60V} have found that about 25\% of the analyzed quiet Sun Stokes $V$ profiles have complex shapes. Even though the authors have not quantified the area or amplitude asymmetry of these complex profiles, they have found, using MIcro-Structured Magnetized Atmosphere \citep[MISMA,][]{1996ApJ...466..537S} inversions, that  these complex/asymmetric profiles belong to pixels in which opposite polarities coexist within the resolution element (angular resolution 0.3\arcsec). They report that these pixels are often observed along the polarity inversion line (PIL) between opposite polarity patches \citep[see also,][]{2001ApJ...563.1031S} or very quiet regions (i.e. regions with weak polarization signals).

\cite{2014ApJ...793L...9K} showed that asymmetric Stokes $V$ profiles along the PIL can be reproduced using a linear combination of Stokes $V$ profiles from either side of the PIL, for a sub-granular scale cancelation event. They suggested that the observed asymmetry is not related to magnetic flux cancelation, and that they are representative of either coexisting opposite polarities or unresolved width of the PIL.

In this paper we focus on quantifying the area asymmetry of Stokes $V$ profiles and its evolution in pixels neighboring the PIL that belong to positive and negative polarity features (hereafter we refer to these features as positive and negative patches). Our aim is to characterize the evolution of the area asymmetry related to reconnection and magnetic flux cancelation. We will also show the stratification with optical depth of the physical parameters at these locations under the assumption of Local Thermodynamic Equilibrium (LTE) using the Stokes Inversion based on Response functions \citep[SIR;][]{1992ApJ...398..375R} code. 

\section{Observations and analysis} \label{sec:obsv}
\begin{figure}
\centering
  \includegraphics[trim=50 325 180 225,clip,width=0.55\textwidth]{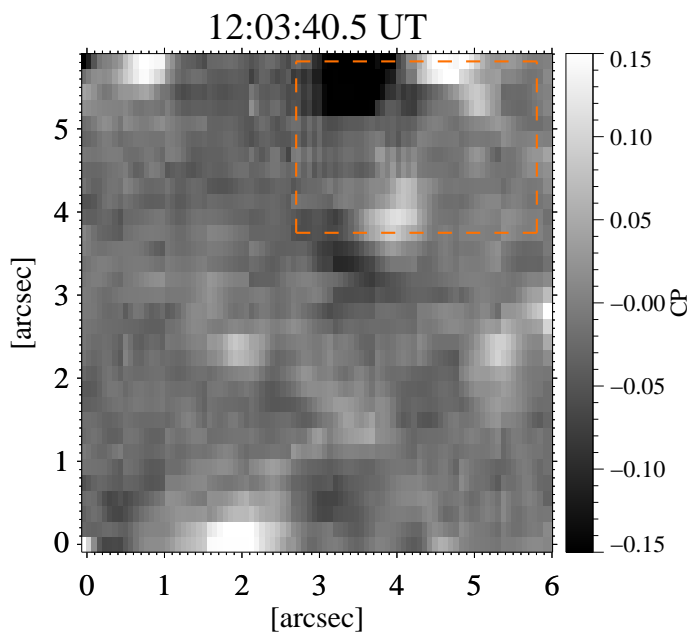}
 \caption{Map of circular polarization (CP). The map is saturated at $\pm$0.15. The red dashed rectangular box encloses our region of interest (ROI), where opposite polarity magnetic features are interacting.}
  \label{fig:a}
\end{figure}

Spectropolarimetric maps of the quiet Sun (see Fig.\ref{fig:a}) at disk centre were collected using the Integral Field Unit (IFU, Dominguez-Tagle et al., in preparation) mounted on the GREGOR Infrared Spectrograph \citep[GRIS; ][]{2012AN....333..872C}, which is attached to the 1.5 m GREGOR telescope \citep{2012AN....333..796S}. The GRIS sampled the photospheric \ion{Si}{I} 10827.108~{\AA} line (effective Land\'{e} factor $g$ = 1.5) with a spectral resolution of 18.0 m{\AA} pix$^{-1}$. The spectral window covered the \ion{He}{I} triplet at 10830~{\AA} as well. However, Stokes $I$ did not show any absorption at this location; therefore, the spectral line \ion{He}{I} at 10830~{\AA} is not present and we cannot use it in our study. The data were taken on 2018 November 02 between 11:35:02 and 12:15:08 UT with a cadence of 26.4~s. The IFU offers a spatial sampling of 0.135\arcsec $\times$ 0.1875\arcsec and covered a field of view (FOV) of 6.075\arcsec $\times$ 6.0\arcsec. Estimating the spatial resolution of the data from the power spectrum of the granular continuum intensity is not an easy task and is unreliable owing to the small FOV of the IFU. Hence, we rely on the root-mean-square of the continuum contrast ($\Delta I_\mathrm{r.m.s.}$) and the seeing parameter ($r_\mathrm{0}$) as an approximate measure of the spatial resolution of the data. The $r_\mathrm{0}$ value during the observation is 22 -- 25cm at 1100 nm, and we obtained a $\Delta I_\mathrm{r.m.s.}$ of 3.3\% at this wavelength. The $\Delta I_\mathrm{r.m.s.}$ value of the Hinode SP at 630 nm is 7\% \citep{2008A&A...484L..17D}.

After calibration \citep{1999ASPC..184....3C, 2003SPIE.4843...55C}, the noise in Stokes $I$ is about 5.3 $\times$ 10$^{-3} <I_\mathrm{QS}>$, where $<I_\mathrm{QS}>$ is the mean quiet Sun continuum intensity. The noise level {\bf ($\sigma$)} in Stokes $Q$, $U$, and $V$ is determined using the polarization signals in the continuum wavelength points and their respective values, in units of $<I_\mathrm{QS}>$, are 9 $\times$ 10$^{-4}$, 1.2 $\times$10$^{-3}$, and 9.4 $\times$ 10$^{-4}$. 

We used the Fourier Transform Spectrometer \citep[FTS, ][]{1999SoPh..184..421N} atlas in the 10830~{\AA} region for the absolute wavelength calibration. The absolute wavelength is then used to determine the LOS velocity ($v_\mathrm{LOS}$) from the Stokes $I$ line core position. This velocity is used only to do a comparison with the $v_\mathrm{LOS}$ value derived from the Stokes $V$ zero-crossing position. Later, $v_\mathrm{LOS}$ will be derived through the Stokes inversions. The five-minute oscillations were removed from the $v_\mathrm{LOS}$, continuum, and line core maps using a subsonic filter \citep{1989ApJ...336..475T} with a cut-off phase velocity of 4 km~s$^{-1}$. We defined circular polarization (CP) as, CP = $V_\mathrm{b} - V_\mathrm{r}$, where $V_\mathrm{b}$, $V_\mathrm{r}$ are the integral of the Stokes $V$ signal, normalized to the mean quiet Sun continuum in the blue and red lobe respectively. We also determined the area asymmetry ($\delta A$) of the Stokes $V$ profiles, which is defined as

\begin{ceqn}
\begin{align*}
\delta A = \frac{\int_{\lambda_i}^{\lambda_f} V(\lambda)~d\lambda} {\int_{\lambda_i}^{\lambda_f} |V(\lambda)|~d\lambda}
\end{align*}
\end{ceqn}

\noindent where $\lambda_i$ = 10825.351~{\AA} and $\lambda_f$ = 10828.952~{\AA}. Stokes $V$ reaches the continuum at both $\lambda_i$ and $\lambda_f$. $\delta A$ takes the sign of the Stokes $V$ blue lobe. We determined $\delta A$ only for those pixels with a maximum Stokes $V$ amplitude above 5$\sigma_\mathrm{v}$, where $\sigma_\mathrm{v}$ is the noise in Stokes $V$. In the entire time series, about 5\% of the pixels satisfy this criterion. We chose a threshold of 5$\sigma_{v}$ to avoid the influence of noise through the following method. First a histogram of Stokes $V$ maximum signal was obtained. We then created a histogram of the noise obtained for each pixel and multiplied the values by $\sigma_\mathrm{v}$. When comparing the Stokes $V$ histogram with the obtained noise histogram, it becomes clear that, at 5$\sigma_\mathrm{v}$ the Stokes $V$ signals are above the highest values of the noise distribution. We found from the resulting area asymmetry maps that about 30\% of the chosen pixels have an area asymmetry above 20\%.

To get an initial impression of the physical parameters in all the pixels inside the region of interest (ROI), we first performed an inversion of the observed Stokes vector using the SIR code with very few free parameters. We considered a single component atmosphere within a resolution element for both, pixels showing a Stokes $V$ area asymmetry and pixels with regular Stokes $V$ profiles. The synthetic profiles are convolved with a spectral point spread function (PSF) of GREGOR/GRIS. For this we used a Gaussian profile with an FWHM of 75 m{\AA}, spectral sampling of 18 m{\AA}, which represents the 10830~{\AA} region, and a spectral stray light correction of 13.5\% \citep[see][for a description of the method of determining and removing the spectral stray light]{2016A&A...596A...2B}. The inversion runs with three nodes in temperature, one in magnetic field strength and inclination, and $v_\mathrm{LOS}$.  As $Q$ and $U$ are mostly below the noise within the ROI (red box in Fig.\ref{fig:a}) we do not consider the azimuthal angle of the magnetic field for inversion, hence no nodes in the azimuth angle. 
\section{Results and Discussion} \label{sec:results}
\begin{figure*}
\centering
  \includegraphics[trim=25 305 90 320,clip,width=0.7\textwidth]{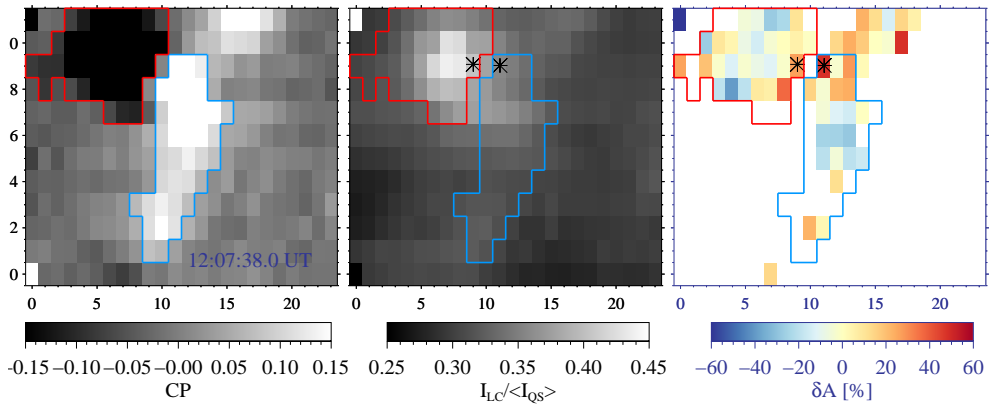}
\caption{A snapshot at 32.6 minutes. From left to right: maps of CP, line core intensity, $I_\mathrm{LC}$, normalized to mean quiet Sun continuum, and area asymmetry. The white pixels in the area asymmetry map represent pixels where the maximum amplitude of Stokes $V$ is below  5~$\sigma_\mathrm{v}$. The asterisk symbols in the middle and the right panels represent pixels from which the profiles in Fig.\ref{fig:d} are taken. The red (blue) contour is for the negative (positive) patch. The axes are in pixel scale (scaling: 0.135\arcsec $\times$ 0.1875\arcsec). A movie of the time series is available online.}
  \label{fig:b}
\end{figure*}

\begin{figure}
\centering
  \includegraphics[trim=30 110 40 90,clip,width=0.5\textwidth]{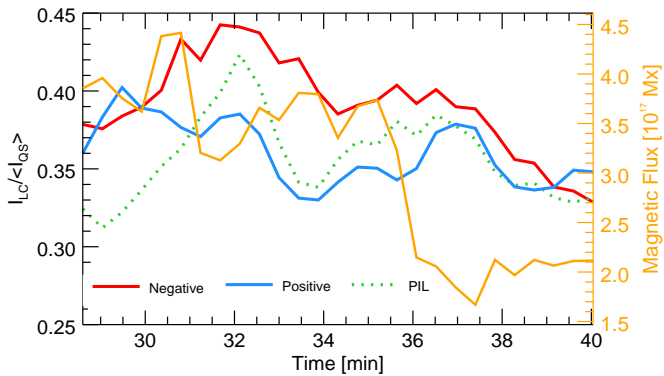}
  \caption{Temporal evolution of the total magnetic flux of the negative patch (orange), and peak normalized line core intensity within the positive (blue) and negative (red) patches, and along the PIL (green).}
  \label{fig:c}
\end{figure}

Figure \ref{fig:b} shows several observed parameters in a single snapshot during the cancelation event in the ROI. They are the circular polarization CP (left panel), the line core intensity normalized to mean quiet Sun continuum $I_\mathrm{LC}$/$<I_\mathrm{QS}>$(middle) and the area asymmetry $\delta A$ (right). The canceling pair (seen in contours in the left panel of Fig.\ref{fig:b}) was visible for a period of about 12 minutes. The contours in Fig.\ref{fig:b} enclose pixels with signal above twice the noise level in the CP maps. The negative patch (red contour) stays compact over time, whereas the positive patch (blue contour) undergoes splitting and merging a number of times (see the online movie). Both features have apparent sizes of sub-granular scale. The positive patch is smaller in size  during the first 8.4 minutes after the initial patch identification. We see a decreasing trend in the total magnetic flux of the negative patch (see Fig.\ref{fig:c}), starting at about 35.5 minutes. The magnetic flux decay rate is $-1.6 \times 10^{15}$ Mx s$^{-1}$. 

\begin{figure}
\centering
  \includegraphics[trim=15 45 30 250,clip,width=0.45\textwidth]{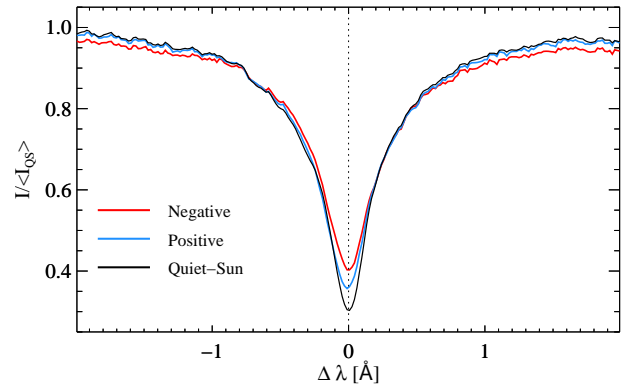}
  \caption{Stokes $I$ profiles, at 32.6 minutes for negative (red) and positive (blue) pixels marked with an asterisk in Fig.\ref{fig:b}. The black line represents the spatio-temporal mean quiet Sun profile. The vertical dotted line indicates the rest centre wavelength. All the three profiles are normalized to mean quiet Sun continuum $<I_\mathrm{QS}>$}.
  \label{fig:d}
\end{figure}

The line core intensity is enhanced within the negative patch compared to the positive patch (see the middle panel of Fig.\ref{fig:b}), with both the positive and negative patches featuring larger line core intensities than the average quiet Sun. This can be confirmed from Fig.\ref{fig:d}, which shows the Stokes $I$ profiles, at 32.6 minutes, of the border pixels marked with asterisks in the last two panels of Fig.\ref{fig:b}, along with the mean quiet Sun profile. All the three profiles are normalized to the mean quiet Sun continuum $<I_\mathrm{QS}>$. The line core intensity of the negative polarity pixel is higher than that of the positive polarity pixel and the quiet Sun value by 13\% and 33\%, respectively. For the positive polarity pixel, the line core intensity is higher by 17.8\% than the mean quiet Sun value. The difference in line core intensity between the negative polarity pixel, the positive polarity pixel, and the quiet Sun is significant as it is much higher than the photon noise. The Doppler velocity of the positive polarity pixel from the Stokes $I$ line core position is $-$0.27 kms$^{-1}$, which is lower than that from the zero-crossing position of Stokes $V$ ($-$1.5 kms$^{-1}$). For the negative polarity pixel, the Doppler velocity from Stokes $I$ and Stokes $V$ profiles are about 0.03 kms$^{-1}$ and 0.3 kms$^{-1}$, respectively.

We are able to relate the line core intensity increase to the ongoing flux decay from its temporal evolution (see Fig.\ref{fig:c}). We see that the line core intensity within the negative patch peaks ($T=31.7$ minutes) before its magnetic flux starts declining ($T\sim 35.5$ minutes). As we do not see at this time an increase in magnetic field strength, we rule out a bright point development due to a shift in the Wilson depression. This event could then be pointing to magnetic flux cancelation resulting from magnetic reconnection above the solar surface. We think that the decrease in magnetic flux around 31.5 minutes results from an abrupt change in the magnetic field configuration of the negative polarity patch due to field line reconnection. The intensity of the PIL peaks 26.4 s later than that of the negative patch. The intensity within the positive patch stays more or less the same during the event. It could be that the reconnection site is spatially closer to the negative patch than the positive and hence the intensity enhancement is greater in the negative patch. 

\begin{figure}
\centering
\subfloat[]{%
  \includegraphics[trim=0 44 0 250,clip,width=0.5\textwidth]{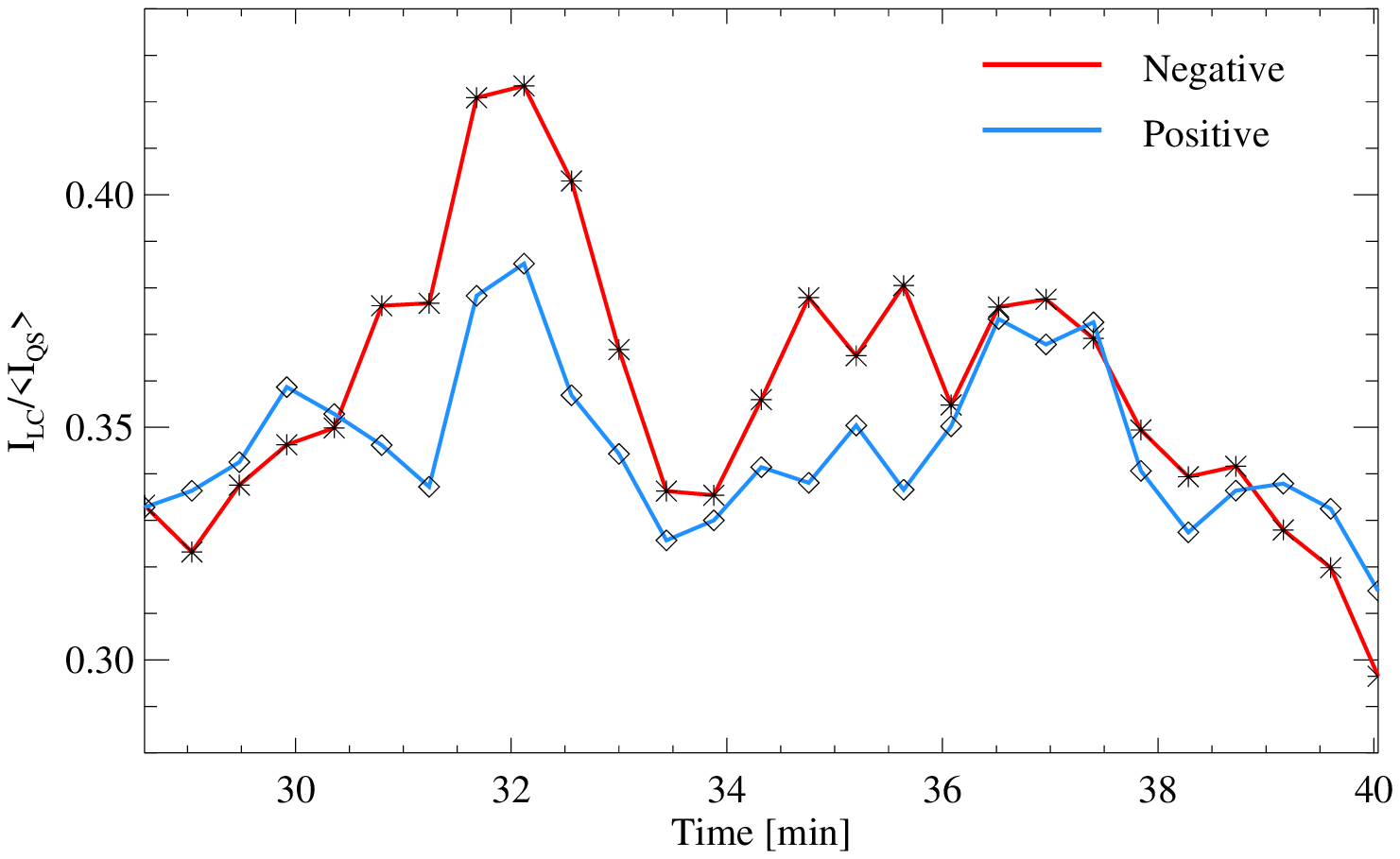}%
  }\par
\subfloat[]{%
  \includegraphics[trim=3 40 0 250,clip,width=0.5\textwidth]{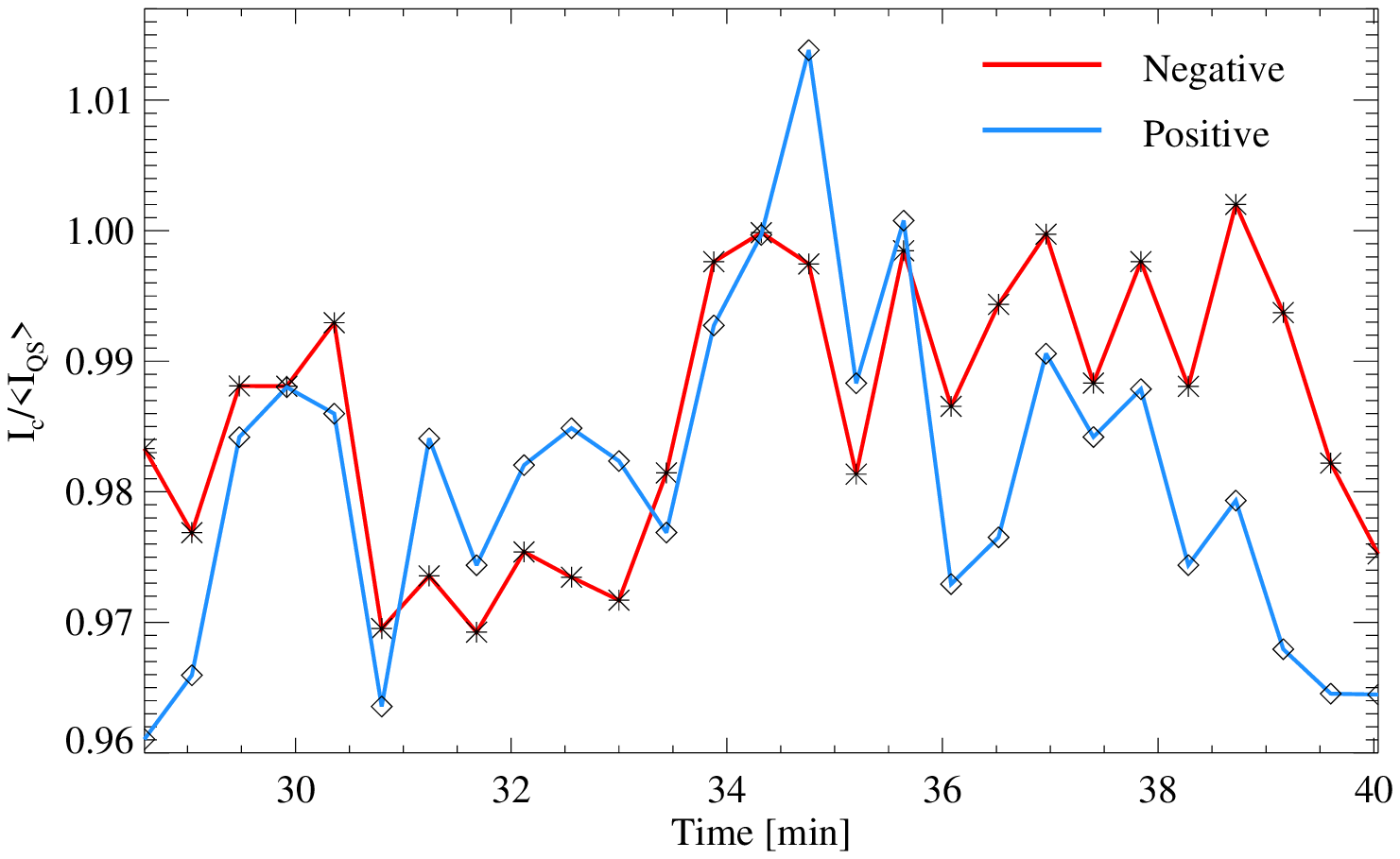}%
  }\par        
\caption{Variation of intensity at the line core (a) and in the continuum (b) for border pixels of both polarities.}
\label{fig:e}
\end{figure}

Figure \ref{fig:c} shows that the line core intensity within the positive patch and along the PIL has two peaks. This trend is not visible for the negative patch. To check whether a similar trait is shown by the border pixels, we made plots of the evolution of intensity at the line core (Fig.\ref{fig:e} (a)) and in the continuum (Fig.\ref{fig:e} (b)), for border pixels of both polarities. From panel (a) we see that there is a strong peak at 32.1 minutes for both polarities. And a relatively weak peak at 35.6 minutes for the negative polarity, and at 37.4 minutes for the positive polarity. Panel (b) shows that the continuum of the negative polarity has a weak peak at 30.4 minutes and a relatively stronger and extended peak starting at 34.3 minutes. Similarly, the positive patch has a weak peak at 29.9 minutes and a stronger peak at 34.7 minutes. From a cross-correlation analysis between the core and the continuum time series, we find that the line core peak at 32.1 minutes corresponds to the continuum peak at 34.3 minutes for the negative polarity, and at 34.7 minutes  for the positive polarity. Considering the strong peaks in the core and the continuum, and assuming that the mean line core formation height is about 400 km \citep[see Fig. 3 in][]{2017A&A...603A..98S}, we found that the energy released, from reconnection, in the upper photosphere travels down to the lower photosphere at a speed of about 3 kms$^{-1}$. 

\begin{figure*}
\centering
 \includegraphics[trim=3 130 10 35,clip,width=0.8\textwidth]{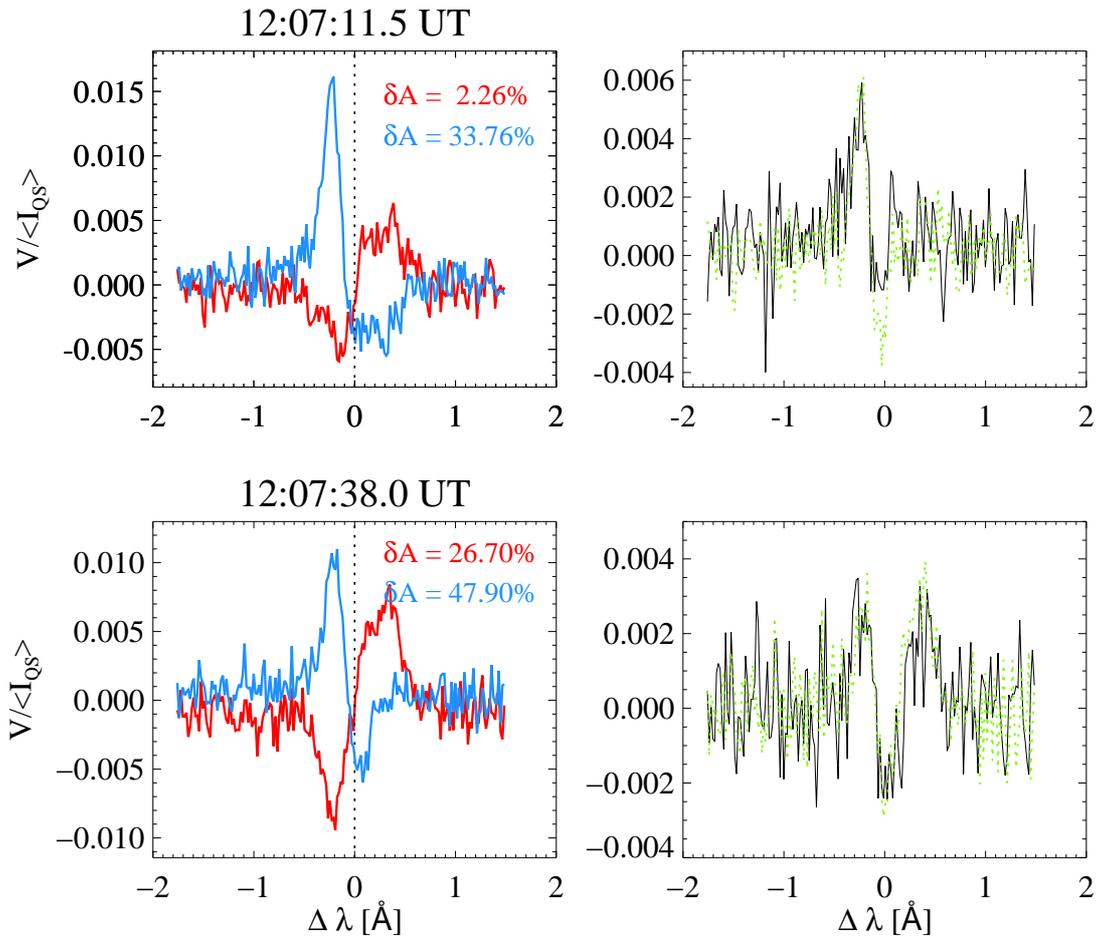}
 \caption{Lower-left panel: Stokes $V$ profiles for pixels marked with asterisks in Fig.\ref{fig:b}. Blue (red) represents positive (negative) polarity pixel. Lower-right: Stokes $V$ profile at the PIL pixel between the asterisk symbols (black). Over-plotted  in green is a synthetic profile created from a linear combination of positive and negative polarity profiles shown in the lower-left panel. The upper panels show same parameters, but for a preceding time frame. Vertical dotted lines on the top and lower-left panels denote the rest centre wavelength.}
\label{fig:f}
\end{figure*}

From the area asymmetry maps (e.g. \ see last panel of Fig.\ref{fig:b}) we find that both polarities have significant $\delta A$ values. A visual confirmation is provided in Fig.\ref{fig:f} -- the lower-left panel shows Stokes $V$ profiles from pixels marked with asterisks in the last two panels of Fig.\ref{fig:b}, and the one on the upper-left panel comes from the preceding frame. In the upper-right and lower-right panels we show in black the Stokes $V$ profiles from the PIL pixel between the asterisk symbols for the respective time frames. In these right-panels we over-plotted a synthetic profile (green dotted line) which is a linear combination of the negative and positive polarity Stokes $V$ profiles shown in the respective left panels. It can be seen that the similarities are quite large. For the upper-right panel, the coefficients of linear combination for the negative and positive polarity Stokes $V$ profiles are 0.53 and 0.47, respectively. And those for the lower-right panel are 0.41 and 0.59, respectively, for the negative and positive polarity Stokes $V$ profiles. This simple test implies that the PIL profiles are spatially unresolved in our data. A similar result was found by \cite{2014ApJ...793L...9K}. 

\begin{figure*}
\centering
\subfloat[]{%
  \includegraphics[trim=10 270 0 365,clip,width=0.8\textwidth]{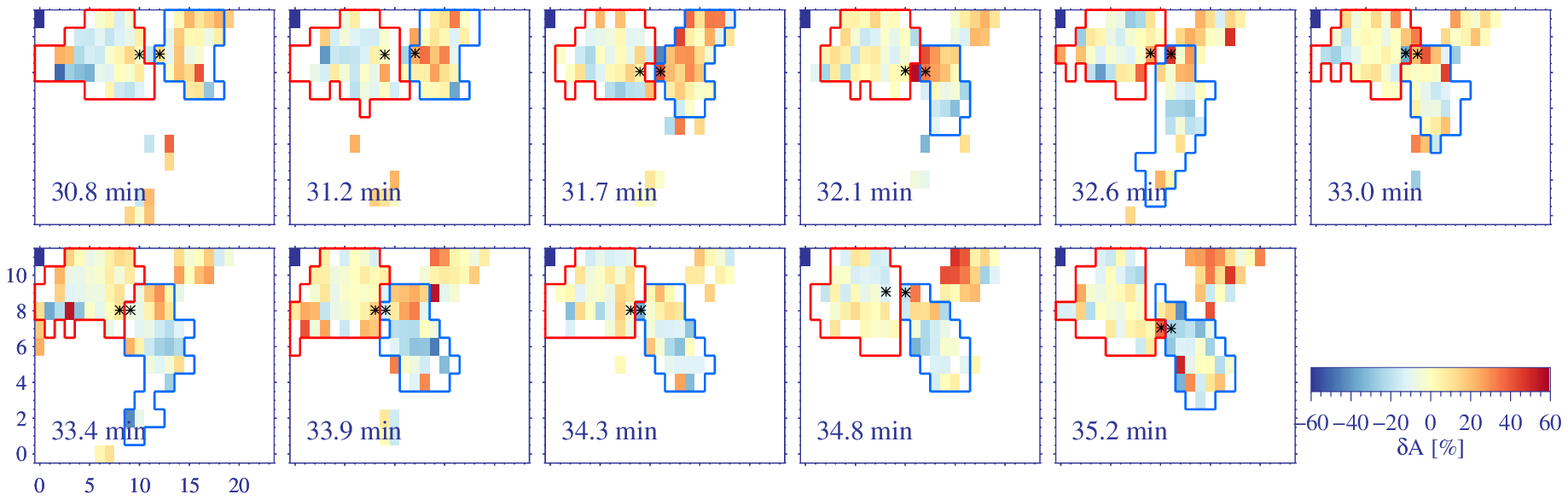}%
  }\par
\subfloat[]{%
  \includegraphics[trim=3 110 60 96,clip,width=0.8\textwidth]{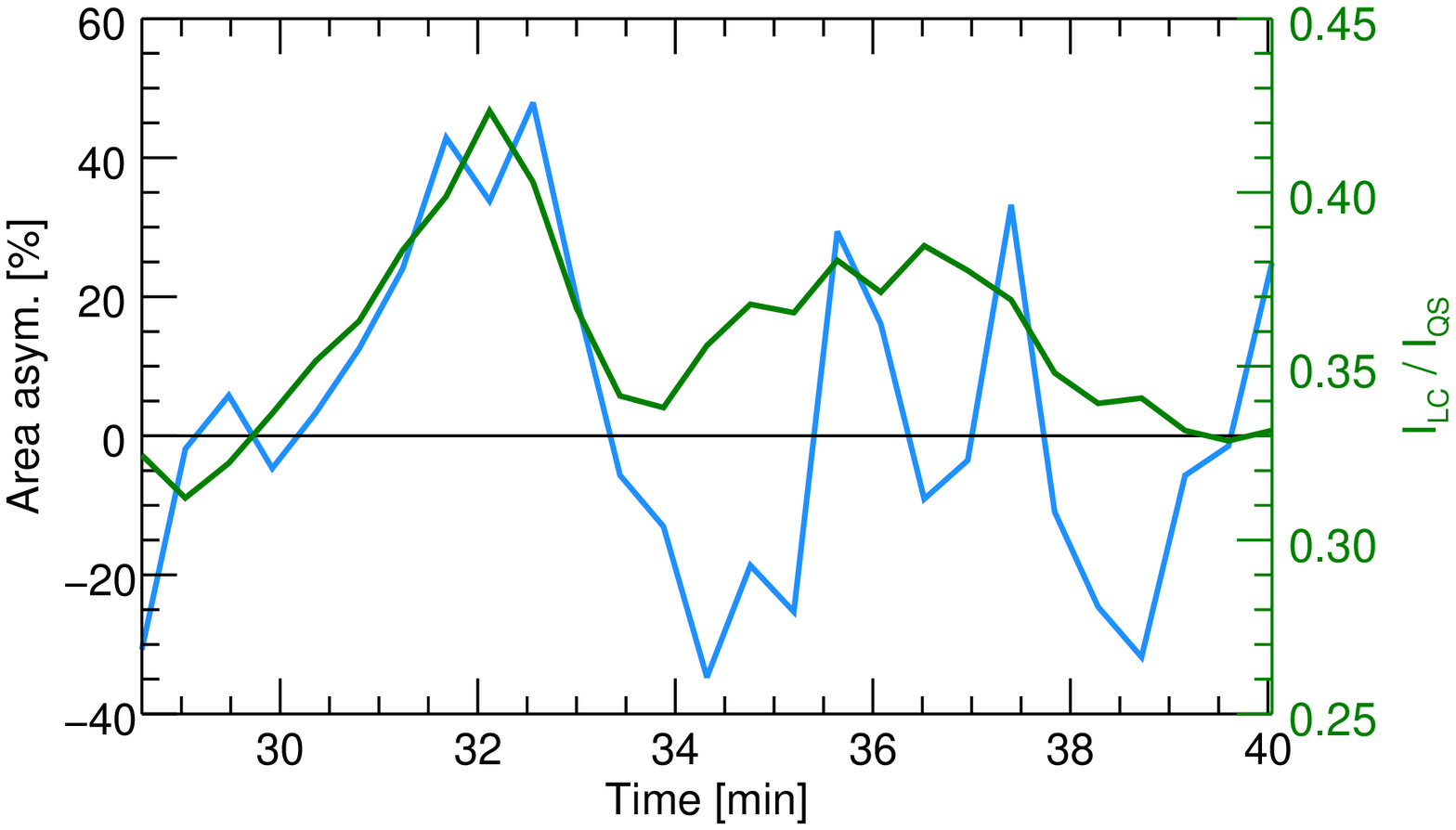}%
  }\par        
\caption{(a) Area asymmetry maps in the interval [30.8, 35.2] minutes. The axes are in pixel scale (scaling: 0.135\arcsec $\times$ 0.1875\arcsec). The red (blue) contour is for the negative (positive) patch. Asterisks represent border pixels whose area asymmetry values are plotted in panel (b). (b) Variation of area asymmetry values for border positive polarity pixels next to the PIL. The evolution of normalized line core intensity along the PIL is shown in green. The horizontal line represents zero on the area asymmetry axis.}
\label{fig:g}
\end{figure*}

To examine whether the area asymmetry of pixels on either side of the PIL (for example, see panel (a) of Fig.\ref{fig:g}) are related to reconnection and flux cancelation, we made a temporal evolution plot of $\delta A$ values from those pixels (the chosen border pixels are marked with asterisks symbols in the online movie), as shown in panel (b) of Fig.\ref{fig:g}. The $\delta A$ values of the positive patch shows a systematic variation over time. It is clear that the area asymmetry amplitude of the positive patch reaches maximum before the onset of cancelation. As the total magnetic flux of the negative patch and the core intensity along the PIL and within the negative patch starts decreasing, the area asymmetry amplitude of the positive patch also decreases. As $\delta A$ is produced by gradients with optical depth in $v_\mathrm{LOS}$ and magnetic field we can conclude that those gradients are larger in the positive patch. And it seems that those gradients relax after the field line reconnection and continue so during flux cancelation. It appears that the circular polarization signals and their area asymmetry at and around the PIL could provide additional information on the reconnection-related cancelation event.

Figure \ref{fig:d} already demonstrated that the temperature in the high photospheric layers at either side of the PIL, as indicated by the line core intensity, is enhanced with respect to the quiet Sun. To quantify this enhancement, we performed additional SIR inversions on the pixels marked with asterisks in Fig.\ref{fig:b}. The frames in this figure correspond to the time when the area asymmetry of the positive patch peaks (32.6 minutes). The Stokes $I$ profiles of the asterisk-labelled pixels are shown in Fig.\ref{fig:d}. To ensure the uniqueness of the inversion results, the pixels are inverted with a thousand different initial atmospheric models. These models are obtained by randomly perturbing the magnetic field parameters, $v_\mathrm{LOS}$, and temperature stratification of the Harvard-Smithonian Reference Atmosphere (HSRA) model for the quiet Sun. Of the thousand runs, we chose only those solutions with smallest $\chi^2$ values, i.e. $\chi^2$ values below the sum of mean $\chi^2$ and the standard deviation. We used four nodes in temperature for both positive and negative polarity pixels. For the magnetic field parameters and $v_\mathrm{LOS}$, we considered two nodes for the positive polarity pixel and one node (i.e. height independent) for the negative polarity pixel, as the area asymmetry is small and likely heavily affected by photon noise. The weights given for the Stokes $I$, $Q$, $U$, and $V$ profiles are 2,1,1, and 4, respectively. Section \ref{sec:obsv} gives details of the spectral PSF profile. The inversion takes into account the contribution of stray light to the Stokes $I$ profile. We took the spatio-temporal average of Stokes $I$ as the stray light profile. 

\begin{figure*}
\centering
 \includegraphics[trim=3 140 5 25,clip,width=0.9\textwidth]{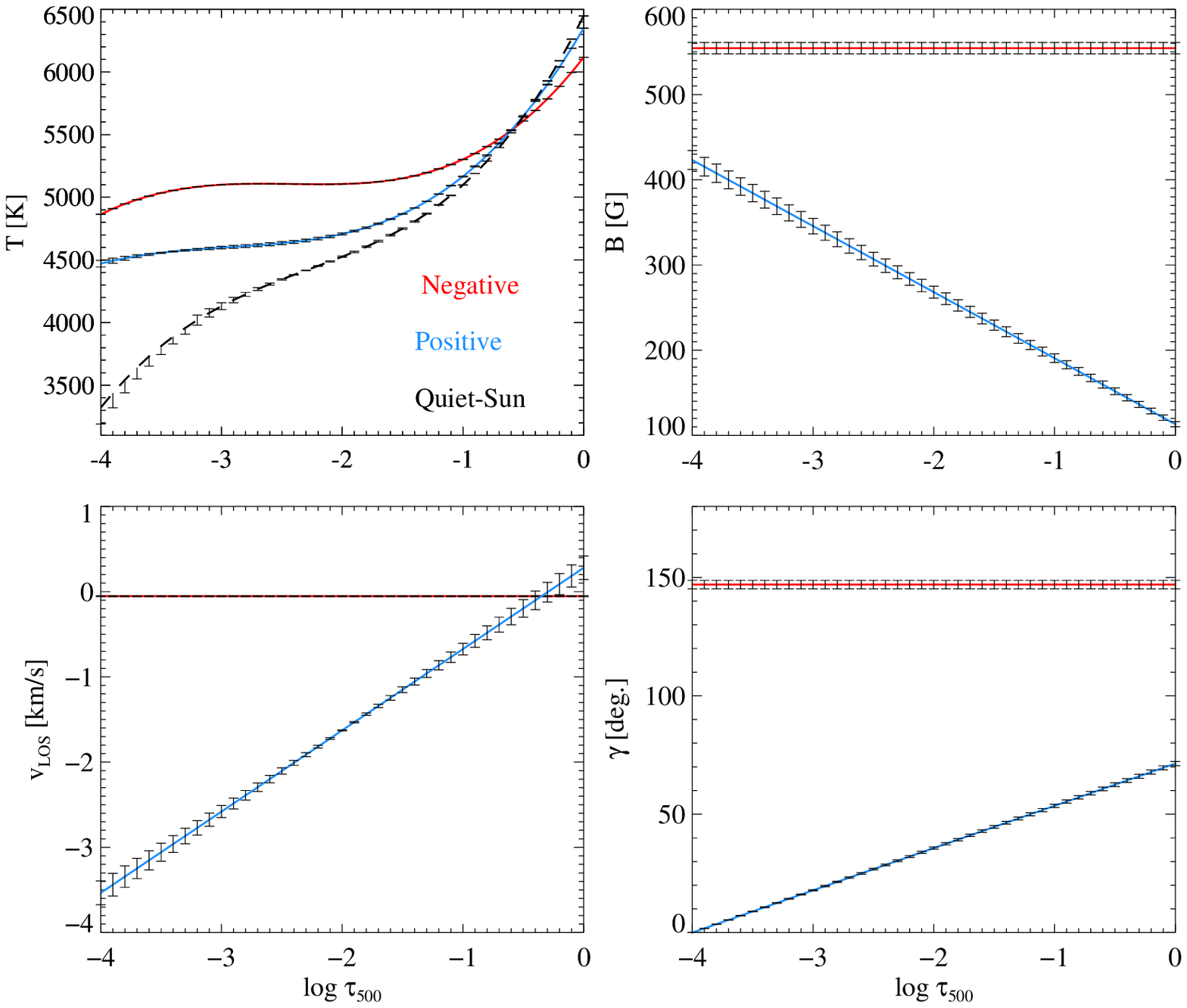}
  \caption{Mean values of physical parameters over the chosen SIR inversion runs on the pixels marked with an asterisk in the last panel of Fig.\ref{fig:b}. For all the panels blue and red represent pixels with positive and negative polarity signals, respectively. At top-left is the excitation temperature variation of positive and negative polarity pixels, and of the mean quiet Sun profile (black dashed line). Top-right: variation of magnetic field. Bottom-right: magnetic field inclination, $\gamma$, with respect to the LOS. Bottom-left: $v_\mathrm{LOS}$. Positive $v_\mathrm{LOS}$ values represent redshifts and negative values represent blueshifts. The error bars represent standard deviation values.} 
\label{fig:h}
\end{figure*}

The resulting excitation temperature, field strength, inclination ($\gamma$), and $v_\mathrm{LOS}$ as a function of optical depth are shown in Fig.\ref{fig:h}. The mean values of the physical parameters over the chosen simulation runs (i.e. \ those with sufficiently small $\chi^2$ values) are plotted. A major contribution to the core of the \ion{Si}{I} line comes from a mean optical depth of log~$\tau_\mathrm{500} = -3$ (upper photosphere) \citep[see Fig.2 in][]{2017A&A...603A..98S}. At this height, the excitation temperature in the the atmosphere of the negative polarity pixel is higher than in the atmosphere of the positive polarity pixel and in the average quiet Sun atmosphere by 502 K and 969 K, respectively (see upper-left panel of Fig.\ref{fig:h}). The excitation temperature of the positive polarity pixel atmosphere is 467 K higher than in the average quiet Sun. For accurate estimation of the kinetic temperature enhancements in the upper photosphere an NLTE inversion should be performed.

The line cores of the border pixels during the steep decline in negative flux also show a similar behaviour, as shown in Fig.\ref{fig:d}. This implies that the temperature enhancement is sustained during cancelation. Such enhancement during small-scale cancelation events happening in the vicinity of an emerging active region was previously reported in \cite{2019A&A...622A.200K}.  Another study by \cite{2005ApJ...626L.125B}, however, found no significant temperature enhancement in either polarity during cancelation occurring in the moat of a sunspot.

In the upper photosphere, log~$\tau_\mathrm{500} < -3$, we expect the quiet Sun temperature stratification returned by SIR to differ significantly from standard 1D models \citep[e.g. MACKKL 1D;][]{1986ApJ...306..284M} simply because our analysis is done under the assumption of LTE. Deeper down (around log~$\tau_\mathrm{500}\sim~-1.0$), NLTE effects are much smaller, so one could naively expect a much better agreement. This, however, is not the case. We do not have a definite answer to this question. All we state at this point is that the height dependences of the kinetic temperature in all atmospheres considered are different.

From the upper-right panel in Fig.\ref{fig:h}, we see that the magnetic field of the pixel on the positive-polarity side increases with decreasing optical depth. In the deep photosphere, log~$\tau_\mathrm{500} = -1.0$, the field strength of the negative polarity pixel is higher than that of the positive polarity pixel by 364 G. If we look at the inclination at the same height, it can be seen that the magnetic field of the positive polarity pixel is highly inclined compared to that of the negative (lower-right panel Fig.\ref{fig:h}). Despite being highly inclined, the transverse magnetic flux density of the positive polarity pixel, $B_\mathrm{t}$ = $B \times sin \gamma \times (1-\alpha_\mathrm{stray})$, is 46 Mxcm$^{-2}$ at log~$\tau_\mathrm{500} = -1.0$, yielding $Q$ and $U$ signals at the level of the noise. Having Stokes $Q$ and $U$ mostly below the noise, within the ROI, imposes an upper limit on the transverse magnetic flux density ($B_\mathrm{t}$) and therefore also limits the LOS inclination of the magnetic field. Hence, the values returned by the inversions are the upper limit of the values of physical parameters such as $B_\mathrm{t}$, LOS inclination, etc.

Another interesting thing to note here is that $v_\mathrm{LOS}$ of the positive-polarity pixel switches from a redshift in the deep photosphere, $v_\mathrm{LOS} = 0.3$ kms$^{-1}$ at log~$\tau_\mathrm{500} = 0.0$, to a blueshift in the upper photospheric layer, $v_\mathrm{LOS} = -2.6$ kms$^{-1}$ at log~$\tau_\mathrm{500} = -3$. 

Differences in the temperature and velocity between pixels on either side of the PIL (see, for instance, the asterisk in the middle panel of Fig.\ref{fig:b}) would certainly be greater if we had a much better spatial resolution. Also, the better the spatial resolution, the better we can distinguish what happens at the PIL and in its neighborhood. However, the finite spatial resolution corresponds to variations perpendicular to the line-of-sight. These cannot affect the area asymmetry because the area asymmetry is only affected by variations along the line-of-sight. Therefore, we expect our results as a function of $\tau_{\rm 500}$ (Fig.\ref{fig:h}) to remain unchanged even if the spatial resolution increases. This is particularly true for the sign of the derivatives; for instance, if the spatial resolution of the observations increases, the slope of the blue curves for the velocity, inclination, and magnetic field in Fig.\ref{fig:h} might change slightly, but the slope will always have the same sign because this is given by the area asymmetry, which is independent of the spatial resolution.

\begin{figure*}
\centering
 \includegraphics[trim=1 60 100 10,clip,width=0.8\textwidth]{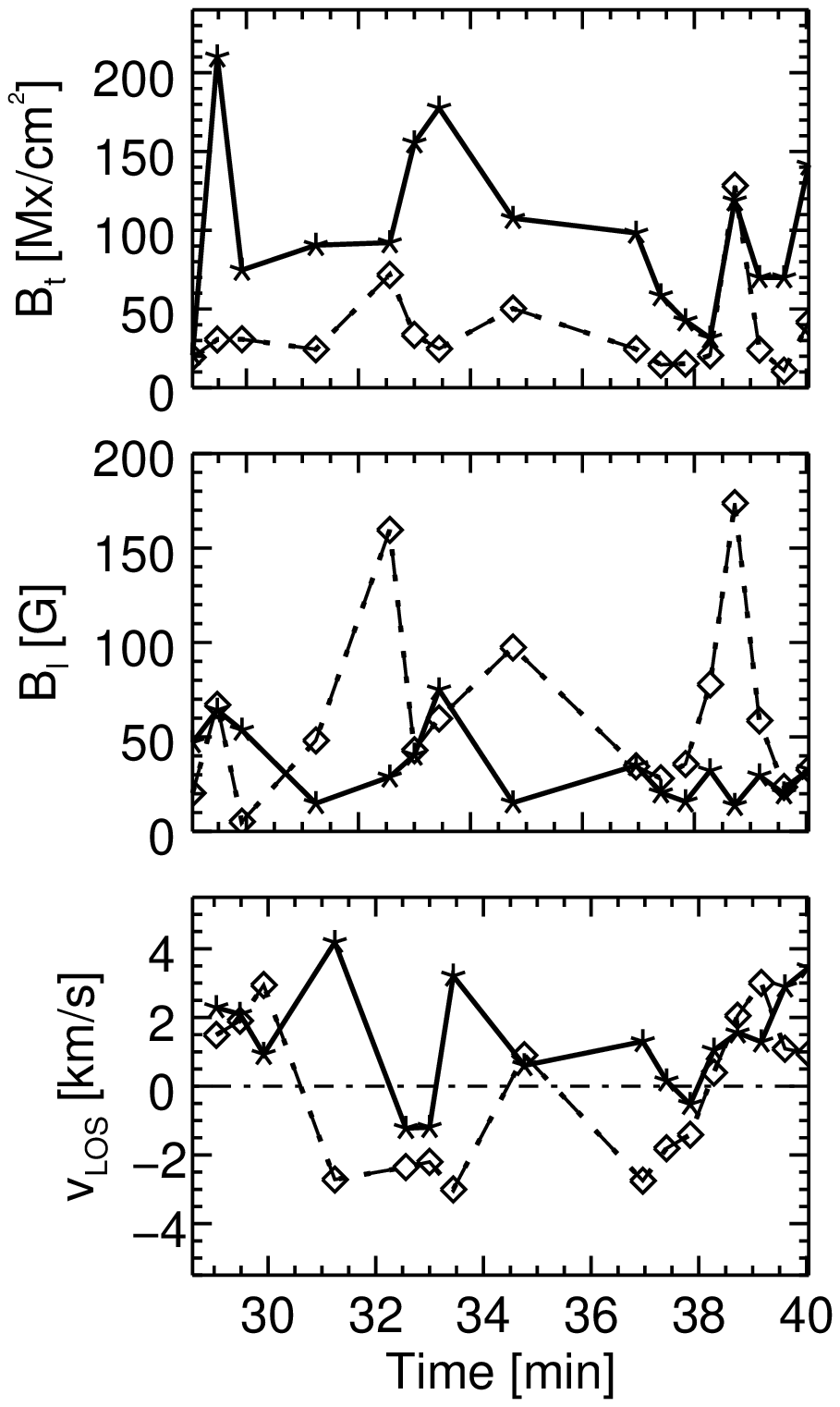}
  \caption{Evolution of transverse magnetic flux density ($B_\mathrm{t}$) (top), longitudinal magnetic flux density ($B_\mathrm{l}$) (middle), and $v_\mathrm{LOS}$ (bottom) for the positive polarity border pixels. Solid lines represent values at  log~$\tau_\mathrm{500} = -1.0$, and the dashed lines correspond to values at log~$\tau_\mathrm{500} = -3$. The black dash-dotted line in the bottom panel denotes zero on the $v_\mathrm{LOS}$ axis.}
  \label{fig:i}
\end{figure*}

To retrieve the temporal evolution of the magnetic field and velocity of the positive polarity border pixels, we inverted those pixels in the time interval [28.6 -- 39.5] minutes, whose area asymmetry values are plotted in Fig.\ref{fig:g} (b), in the same way as that at 32.6 minutes. The only difference here is that we chose a higher weight of 11 for the Stokes $V$ profiles to account for its significant area asymmetry. Fig.\ref{fig:i} shows the variation of transverse magnetic flux density ($B_\mathrm{t}$), longitudinal magnetic flux density ($B_\mathrm{l}$ =  $(1-\alpha_\mathrm{stray}) \times B \times cos \gamma)$, and $v_\mathrm{LOS}$. We found that $B_\mathrm{t}$ (upper panel) values at both optical depth levels, log~$\tau_\mathrm{500} = -3$ and $-1$, are mostly below 200 Mxcm$^{-2}$, except at one point, which is consistent with having the observed Stokes $Q$ and $U$ below the noise level. Values of $B_\mathrm{l}$ (middle panel) are small as well. From the lower panel, it appears that the plasma flow at log~$\tau_\mathrm{500} = -3$ undergoes a cyclic variation. In the case of plasma flow at log~$\tau_\mathrm{500} = -1$, this behaviour is not very evident. 

Figure \ref{fig:g} (b) also shows that the area asymmetry of the positive patch (blue curve) reverses its sign cyclically. This trend is not seen in the area asymmetry of the border pixels of the negative patch. We will investigate in our future work: a) whether this cyclic variation arise from oscillations/waves triggered by the interacting opposite polarities, and, if so, their characteristics and propagation to higher layers; b) whether the sign reversal in area asymmetry is due to a change in sign of the gradient in magnetic field or $v_\mathrm{LOS}$, and c) how we can improve on our analysis with a more accurate determination of the temperature enhancements around the reconnection site by performing NLTE inversions.

\section{Summary} \label{sec:conc} 

Using spectropolarimetric data from the GRIS/IFU, we analyzed a quiet Sun, small-scale cancelation event, at disk centre. This is a reconnection-related cancelation. We observe a significant rise in the line core intensity and temperature prior to cancelation. 

We could reproduce Stokes $V$ profiles along the PIL using a linear combination of asymmetric Stokes $V$ profiles from either side of the PIL. The Stokes $V$ profiles on the border pixels of the positive polarity patch show a systematic variation of area asymmetry, the amplitude of which decreases over time. Border pixels of both polarities show sign reversal in area asymmetry during the event. 

We wish to point out that the SIR inversions we did are based on the LTE approximation of the \ion{Si}{I} 10827~{\AA} line. A correct approach would be to include NLTE effects in the inversion \citep[]{2008ApJ...682.1376B, 2017A&A...603A..98S}, which will be carried out in the future. While it might be true that the amplitude of the profiles changes (owing to deviation of the source function from Planck's function) when considering  NLTE effects, the area asymmetry will be affected to a much lesser extent. Therefore, our results obtained via the analysis of the area asymmetry (for example, the gradients presented in Fig.\ref{fig:h}) will still be present even if an NLTE treatment were to be conducted.

We have analysed only one event here and this may limit our study. To confirm that asymmetric Stokes $V$ profiles on pixels adjacent to the PIL can indeed point to a reconnection-related cancelation, we need more quiet Sun samples, which we intend to analyze in the future.

\begin{acknowledgements}
We thank an anonymous referee for comments that improved the manuscript. The 1.5 m GREGOR solar telescope was built by a German consortium under the leadership of the Kiepenheuer-Institut f\"{u}r Sonnenphysik in Freiburg with the Leibniz-Institut f\"{u}r Astrophysik Potsdam, the Institut f\"{u}r Astrophysik G\"{o}ttingen, and the Max-Planck-Institut f\"{u}r Sonnensystemforschung in G\"{o}ttingen as partners, and with contributions by the Instituto de Astrof\'{i}sica de Canarias and the Astronomical Institute of the Academy of Sciences of the Czech Republic. AJK and CEF are funded by the SAW-2018-KIS-2-QUEST project. CDT and MC acknowledge financial support from the Spanish Ministerio de Ciencia, Innovaci\'on y Universidades through project PGC2018-102108-B-I00 and FEDER funds. We thank M. Franz, H. Strecker and O. Wiloth for their support during the observation campaign. This research has made use of NASA’s Astrophysics Data System.
\end{acknowledgements}

\bibliographystyle{aa}
\bibliography{mybib}
\end{document}